\documentclass[pdftex,twocolumn,epjc3]{svjour3}          % twocolumn

\RequirePackage[T1]{fontenc}

\smartqed  % flush right qed marks, e.g.~at end of proof

\RequirePackage{graphicx}
\RequirePackage{mathptmx}      % use Times fonts if available on your TeX system
\RequirePackage{flushend}
\RequirePackage[colorlinks,citecolor=blue,urlcolor=blue,linkcolor=blue]{hyperref}

\usepackage{amssymb}
\usepackage{mathtools}
\usepackage{amsfonts}
\usepackage{cancel}
\usepackage{cuted}
\usepackage{appendix}
\usepackage{subfig}
\usepackage{graphicx}
\usepackage[labelfont=bf]{caption}
\usepackage{float}
\usepackage{amsmath}
\usepackage{physics}
\usepackage{braket}
\usepackage[
    backend=bibtex,
    maxbibnames=1,
    minbibnames=1,
    giveninits=true,
    sorting=none,
    style=numeric-comp,
    doi=true,
    url=false,
    isbn=false,
    backref=false,
    maxcitenames=2, % adjust as needed
    maxalphanames=1, % adjust as needed
    firstinits=true,
]{biblatex}
\usepackage{url}

\usepackage{caption}
\usepackage{soul}

\usepackage{booktabs}

\usepackage{lineno}
\journalname{Eur. Phys. J. C}

\makeatletter
\renewcommand{\l@section}{\@dottedtocline{1}{0.5em}{3.5em}}
\makeatother

\renewcommand\appendixname{App.}

\begin{document}

\title{Constraints on self-interaction cross-sections of dark matter in universal bound states from direct detection}

\author{
  G.~Angloher\thanksref{addrMPI}\and
  S.~Banik\thanksref{addrHEPHY,addrAI}\and
  G.~Benato\thanksref{addrLNGS,addrGSSI}\and
  A.~Bento\thanksref{addrMPI,addrCoimbra}\and 
  A.~Bertolini\thanksref{addrMPI}\and 
  R.~Breier\thanksref{addrBratislava}\and
  C.~Bucci\thanksref{addrLNGS}\and 
  J.~Burkhart\thanksref{addrHEPHY} \and
  E.~Cipelli\thanksref{addrMPI}\and
  L.~Canonica\thanksref{addrMPI,addrMilan}\and 
  A.~D'Addabbo\thanksref{addrLNGS}\and
  S.~Di~Lorenzo\thanksref{addrMPI}\and
  L.~Einfalt\thanksref{addrHEPHY,addrAI}\and
  A.~Erb\thanksref{addrTUM,addrWMI}\and
  F.~v.~Feilitzsch\thanksref{addrTUM}\and 
  S.~Fichtinger\thanksref{addrHEPHY}\and
  D.~Fuchs\thanksref{addrMPI}\and 
  A.~Garai\thanksref{addrMPI}\and 
  V.M.~Ghete\thanksref{addrHEPHY}\and
  P.~Gorla\thanksref{addrLNGS}\and
  P.V.~Guillaumon\thanksref{addrMPI,addrSaoPaulo}\and
  S.~Gupta\thanksref{addrHEPHY, cor1}\and 
  D.~Hauff\thanksref{addrMPI}\and 
  M.~Ješkovsk\'y\thanksref{addrBratislava}\and
  J.~Jochum\thanksref{addrTUE}\and
  M.~Kaznacheeva\thanksref{addrTUM}\and
  A.~Kinast\thanksref{addrTUM}\and
  S.~Kuckuk\thanksref{addrTUE}\and
  H.~Kluck\thanksref{addrHEPHY}\and
  H.~Kraus\thanksref{addrOxford}\and 
  A.~Langenk\"amper\thanksref{addrMPI}\and 
  M.~Mancuso\thanksref{addrMPI}\and
  L.~Marini\thanksref{addrLNGS}\and
  B.~Mauri\thanksref{addrMPI}\and
  L.~Meyer\thanksref{addrTUE}\and
  V.~Mokina\thanksref{addrHEPHY}\and
  M.~Olmi\thanksref{addrLNGS}\and
  T.~Ortmann\thanksref{addrTUM}\and
  C.~Pagliarone\thanksref{addrLNGS,addrCASS}\and
  L.~Pattavina\thanksref{addrLNGS,addrMilan}\and
  F.~Petricca\thanksref{addrMPI}\and 
  W.~Potzel\thanksref{addrTUM}\and 
  P.~Povinec\thanksref{addrBratislava}\and
  F.~Pr\"obst\thanksref{addrMPI}\and
  F.~Pucci\thanksref{addrMPI}\and 
  F.~Reindl\thanksref{addrHEPHY,addrAI} \and
  J.~Rothe\thanksref{addrTUM}\and 
  K.~Sch\"affner\thanksref{addrMPI}\and 
  J.~Schieck\thanksref{addrHEPHY,addrAI, cor2}\and 
  S.~Sch\"onert\thanksref{addrTUM}\and 
  C.~Schwertner\thanksref{addrHEPHY,addrAI}\and
  M.~Stahlberg\thanksref{addrMPI}\and 
  L.~Stodolsky\thanksref{addrMPI}\and 
  C.~Strandhagen\thanksref{addrTUE}\and
  R.~Strauss\thanksref{addrTUM}\and
  I.~Usherov\thanksref{addrTUE}\and
  F.~Wagner\thanksref{addrHEPHY}\and 
  V.~Wagner\thanksref{addrTUM}\and 
  V.~Zema\thanksref{addrMPI}
  (CRESST Collaboration)
}

\institute
{Max-Planck-Institut f\"ur Physik, D-85748 Garching, Germany \label{addrMPI} \and
Institut f\"ur Hochenergiephysik der \"Osterreichischen Akademie der Wissenschaften, A-1010 Wien, Austria\label{addrHEPHY} \and
Atominstitut, Technische Universit\"at Wien, A-1020 Wien, Austria \label{addrAI} \and
INFN, Laboratori Nazionali del Gran Sasso, I-67100 Assergi, Italy \label{addrLNGS} \and
Comenius University, Faculty of Mathematics, Physics and Informatics, 84248 Bratislava, Slovakia \label{addrBratislava} \and
Physik-Department, TUM School of Natural Sciences, Technische Universit\"at M\"unchen, D-85747 Garching, Germany \label{addrTUM} \and
Eberhard-Karls-Universit\"at T\"ubingen, D-72076 T\"ubingen, Germany \label{addrTUE} \and
Department of Physics, University of Oxford, Oxford OX1 3RH, United Kingdom \label{addrOxford} \and
also at: LIBPhys-UC, Departamento de Fisica, Universidade de Coimbra, P3004 516 Coimbra, Portugal \label{addrCoimbra} \and
also at: Walther-Mei\ss ner-Institut f\"ur Tieftemperaturforschung, D-85748 Garching, Germany \label{addrWMI} \and
also at: GSSI-Gran Sasso Science Institute, I-67100 L'Aquila, Italy \label{addrGSSI} \and
also at: Dipartimento di Ingegneria Civile e Meccanica, Università degli Studi di Cassino e del Lazio Meridionale, I-03043 Cassino, Italy\label{addrCASS} \and
also at: Instituto de Física da Universidade de São Paulo, São Paulo 05508-090, Brazil \label{addrSaoPaulo} \and
also at: Dipartimento di Fisica, Università di Milano Bicocca, Milano, 20126, Italy \label{addrMilan}
}

% \author{
%   F.~Wagner\thanksref{cor, addrHEPHY}\and 
%   F.~Reindl\thanksref{addrHEPHY,addrAI} \and
%   K.~Niedermayer\thanksref{addrHEPHY,addrAI} \and
%   J.~Schieck\thanksref{addrHEPHY,addrAI}
%   (The CRESST Collaboration) \and
%   W.~Waltenberger\thanksref{addrASC,addrCAIML} \and
%   C.~Heitzinger\thanksref{addrASC,addrCAIML} 
%   \FW{more to come in the final version - fork for thesis and publication}
% }

\thankstext{cor1}{e-mail: shubham.gupta@oeaw.ac.at}
\thankstext{cor2}{e-mail: jochen.schieck@oeaw.ac.at}


\maketitle

\begin{abstract}
$\Lambda$- Cold Dark Matter ($\Lambda$CDM) has been successful at explaining the large-scale structures in the universe but faces severe issues on smaller scales when compared to observations. Introducing self-interactions between dark matter particles claims to provide a solution to the small-scale issues in the $\Lambda$CDM simulations while being consistent with the observations at large scales. The existence of the energy region in which these self-interactions between dark matter particles come close to saturating the S-wave unitarity bound can result in the formation of dark matter bound states called darkonium. In this scenario, all the low energy scattering properties are determined by a single parameter, the inverse scattering length $\gamma$. In this work, we set bounds on $\gamma$ by studying the impact of darkonium on the observations at direct detection experiments using data from CRESST-III and XENON1T. The exclusion limits on $\gamma$ are then subsequently converted to exclusion limits on the self-interaction cross-section and compared with the constraints from astrophysics and N-body simulations. 
\end{abstract}

\section{Introduction}\label{sec:intro}

Various observational evidence shows the ubiquitous presence of dark matter (DM) that makes up for around 26.4\% of the mass-energy content of the universe~\cite{planck_2018, Massey_2010, Lelli_2016}. Several theoretical motivations suggest it to be composed of fundamental particles (denoted by $\chi$), the hunt for which has been going on for many decades in direct detection and indirect detection experiments, as well as in colliders.

The $\Lambda$CDM model of the universe is frequently considered as the standard model of Big Bang cosmology, where DM is considered to be cold and collisionless. Earlier numerical simulations performed under the $\Lambda$CDM paradigm without ordinary matter showed remarkable agreement with the observational surveys on large-scale structures~\cite{NFW_1997,sdss_2004}. However, the model faces challenges on small scales, such as the cusp-core problem, the diversity problem, the too-big-to-fail problem, etc.~\cite{blok_2010, Bullock_2017, Kaplinghat_2019, Oman_2015} which have motivated to consider a refinement of the model. It has been shown that introducing the effect of baryonic feedback or self-interactions between DM particles provides a solution to the small-scale issues with $\Lambda$CDM~\cite{TULIN_2018}. The observations require low self-interaction cross-sections at relativistic velocities (on cluster scales) and increased ones at lower velocities (on smaller scales) to account for the observed structures at both scales, implying a velocity-dependence of the self-interaction cross-section~\cite{Randall_2008,Elbert_2015}.

In Ref.~\cite{Braaten_2013}, it is assumed that there exists a velocity range in which these strong self-interactions between the DM particles come close to saturating the S-wave unitarity bound. In this case, all the low energy scattering properties are dependent on one single parameter, i.e., the large scattering length $a$, or equivalently the small inverse scattering length $\gamma$. This assumption requires the existence of an S-wave resonance near the scattering threshold. And, in the case where the resonance is below the scattering threshold, DM can exist as a bound state of two particles $\chi$, called \textit{darkonium} (D) with twice the mass m$_{\chi}$ where the binding energy of the bound state also depends on $\gamma$. The scattering properties of darkonium are also dependent on $\gamma$. Ref.~\cite{Laha_2014} shows that the existence of darkonium would thus impart a different recoil energy spectrum at direct detection experiments compared to the scattering of a single DM particle. The difference can be understood as arising from the extended structure of the incoming particles and the difference in their number density in the solar neighborhood. The reader is referred to \cite{Braaten_2013, Laha_2014, Laha_2015} for more details on the given model. We assume that the darkonium formed late in the universe, i.e., at lower redshifts, and thus, the DM relic density consists of darkonium bound states only, and contributions from single $\chi$ can be neglected.

In this work, we explore the impact of darkonium on the expected recoil spectrum and present the direct detection results on the given theoretical model of the self-interacting dark matter (SIDM), using recent data from CRESST-III~\cite{DetA_2019} and from XENON1T~\cite{Xenon1T_2019}. In Sec.~\ref{sec:Expected_spec}, we discuss the expected recoil spectrum of darkonium as given in Ref.~\cite{Laha_2014} and calculate the spectra for CaWO$_4$ and Xe. The details of the framework used to calculate the physics results are discussed in Sec.~\ref{sec:likelihood}. These results are presented and discussed in Sec.~\ref{sec:Conclusion}.

\section{Expected darkonium recoil spectrum}\label{sec:Expected_spec}

The elastic self-scattering cross-section $\sigma_{\chi-\chi}$ of the DM particles that forms the darkonium has a simple dependence on the relative momentum $k$ between the DM particles~\cite{Laha_2014}:

\begin{equation}\label{eq:sigma_gamma}
\sigma_{\chi-\chi} = \frac{8\pi}{\gamma^2 + k^2}
\end{equation}

with $\textbf{k}=m_\chi \mathbf{v} /2$. Here, we assume that the constituents of darkonium are indistinguishable bosons. If the two constituents are distinguishable fermions, the numerator would be $4\pi$~\cite{Braaten_2013}. 

It is argued in Ref.~\cite{Laha_2014} that the interaction of darkonium with the detector nuclei can result in two possible final states: one being the darkonium interacts elastically with the detector nucleus and remains intact after the collision (elastic scattering scenario), and the second being the bound state breaks apart due to the interaction with the nucleus (break-up scenario). The differential scattering rate expression for both scenarios can be found in Ref.~\cite{Laha_2014}. 

The darkonium elastic self-scattering cross-section $\sigma_{D-D}$ has a more complicated dependence on the relative momentum $k_D$ between the darkonium bound states. For $k_D$ less than or comparable to $|\gamma|$, it can be approximated by: 

\begin{equation}\label{eq:D_D}
\sigma_{D-D} \approx \frac{8\pi}{\gamma_D^2 + k_D^2}
\end{equation}

where $\gamma_D$ is the darkonium inverse scattering length. If the SIDM particles are spin-$\tfrac12$ fermions, $\gamma_D  = 0.6 \gamma$. For identical bosons, $|\gamma_D|$ is almost always greater than $\gamma/3$, although it could be much smaller if $\gamma$ is near the critical values for which there is a four-boson bound state at the two-darkonium threshold~\cite{Laha_2014}. For $k_D$ much larger than $|\gamma|$, the elastic cross section is much smaller than the inclusive cross section (elastic + break-up). The inclusive cross-section is approximately $4\sigma_{\chi-\chi}$ evaluated at $k = k_D/2$ because the darkonium is a loosely bound state of the two DM particles, and either constituent of one darkonium can scatter from either constituent of the other \cite{Braaten_2006}.

As mentioned above, in this framework darkonium only formed at very low redshifts, so the small scale issues are in practice addressed by elastic scatterings that happen between two free DM particles. Consequently, we neglect the impact of darkonium self-interactions and focus solely on the first-order self-interactions among DM particles for further calculation.

\begin{figure*}[t!]
\centering
\includegraphics[width=.48\textwidth, height=0.30\textwidth]{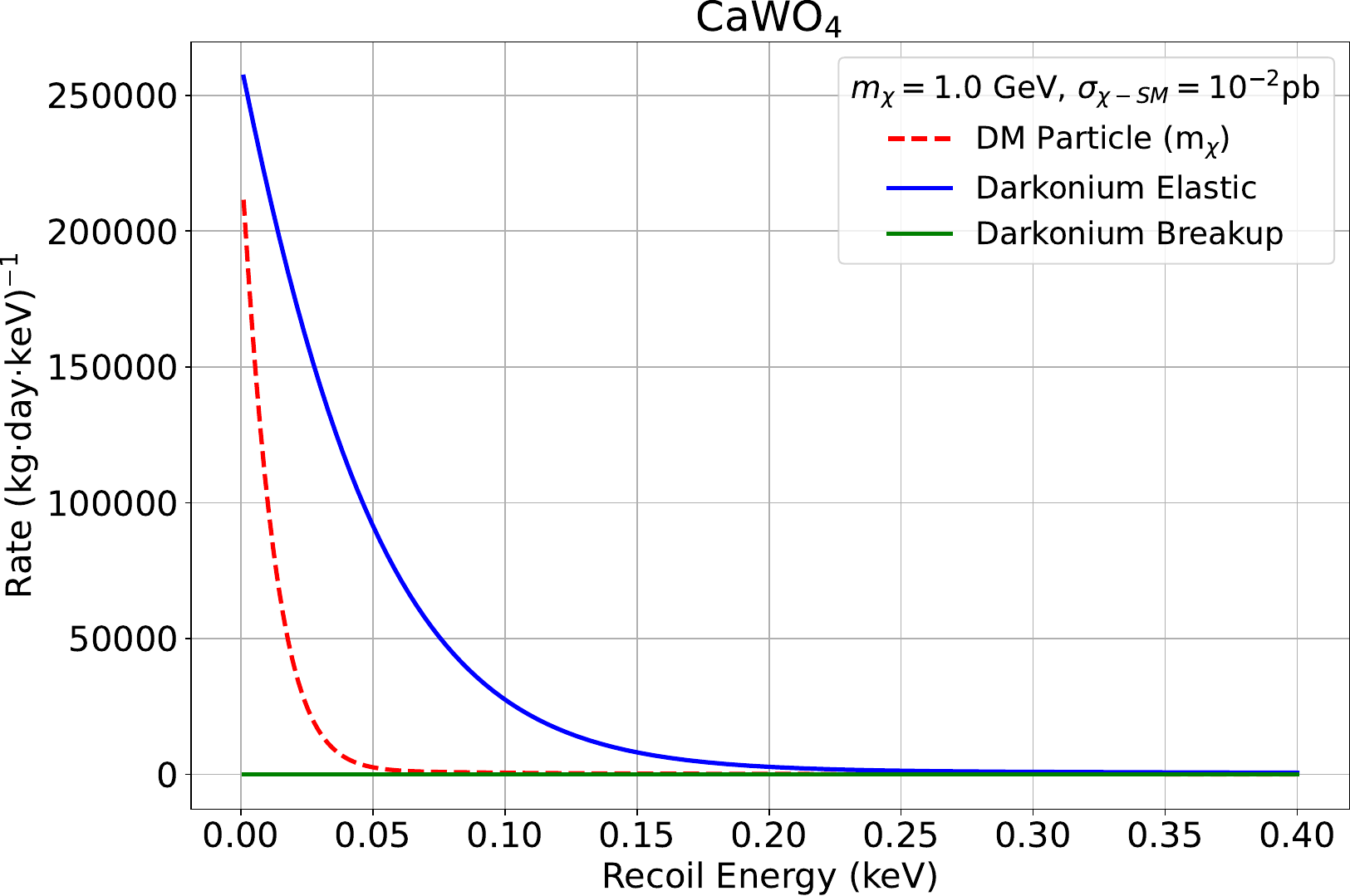}
\vspace{5mm}
\includegraphics[width=.48\textwidth, height=0.30\textwidth]{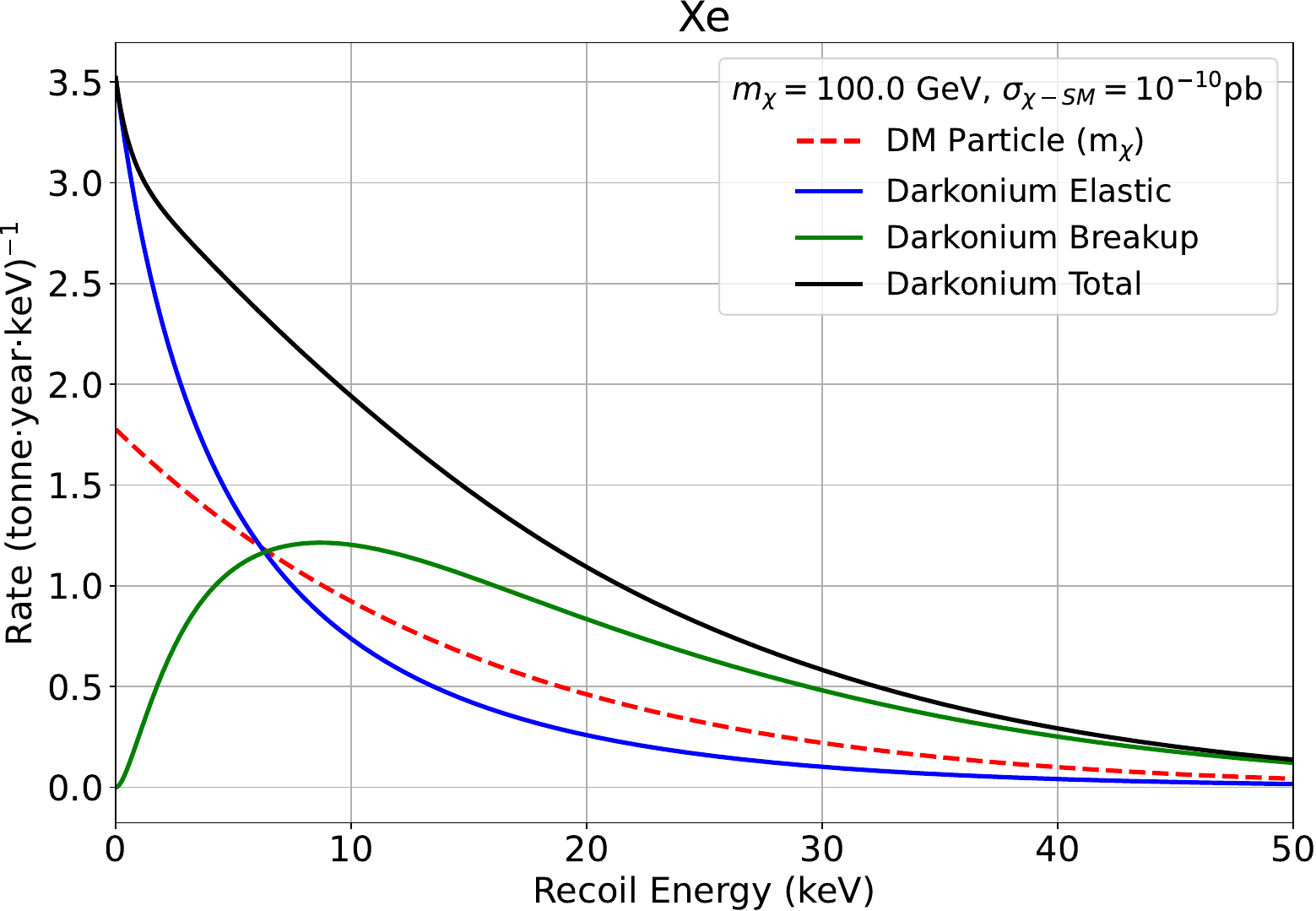}\\
\caption{Expected recoil spectrum for DM particle scattering (red dashed), darkonium elastic scattering (blue solid), darkonium break-up (green solid) and darkonium total scattering (dark blue solid) on CaWO$_4$ nuclei (left) and Xe nuclei (right). The reference cross-section is taken as $\sigma_{\chi-SM}$ = 10$^{-2}$~pb for $m_\chi$ = 1.0 GeV/c$^2$ (left) and $\sigma_{\chi-SM}$ = 10$^{-10}$~pb for $m_\chi$ = 100.0 GeV/c$^2$ (right). The total scattering scenario is only shown for the 100.0~GeV/c$^2$ case, since for 1.0~GeV/c$^2$, it is equivalent to the elastic scattering rate because the break-up scenario is suppressed.}
\label{fig:darkonium_spec}
\end{figure*}

In Fig.~\ref{fig:darkonium_spec}, we show the calculated expected nuclear recoil spectrum for two different DM masses of $m_\chi$ = 1.0 and 100.0 GeV/c$^2$. For the lower DM mass (1.0~GeV/c$^2$), the spectra are calculated in counts per kg$\cdot$day$\cdot$keV and for scattering off CaWO$_4$ nuclei, as this is the material CRESST used in order to set the most stringent spin-independent exclusion limits on the DM-nucleus scattering cross-section at this DM mass~\cite{DetA_2019}. For the higher DM mass (100.0 GeV/c$^2$), the spectra are calculated in counts per tonne$\cdot$year$\cdot$keV for scattering off Xe nuclei, as liquid noble gas experiments take the lead for said mass~\cite{Xenon1T_2019, XenonNT_2023, LZ_2023}. 

The nuclear and astrophysical parameters required to calculate the spectra are the same as those used to calculate the results in Ref.~\cite{DetA_2019}. For each mass, the spin-independent DM-nucleon scattering cross-section ($\sigma_{\chi-SM}$) used to calculate the spectra is just below or at the current exclusion limit for said mass. The value of $\gamma$ is calculated for both masses based on the same description used in Ref.~\cite{Laha_2014}, i.e. using $\sigma_{\chi-\chi}/m_\chi$ = 1~cm$^2$/g at $v=10$~km/s, and solving Eq.~\ref{eq:sigma_gamma}. This value of $\sigma_{\chi-\chi}/m_\chi$ is the typical cross-section required at the dwarf-scale velocities in order to solve the small-scale structure problem in $\Lambda$CDM. The impact of varying $\gamma$ on the nuclear recoil spectrum can be found in Ref.~\cite{gupta_phd}.

As already argued in Ref.~\cite{Laha_2014}, the break-up scenario is suppressed for lower DM masses, which can be seen in the expected spectrum in Fig.~\ref{fig:darkonium_spec}. In the cases considered, the break-up scenario only contributes to the expected rate for the 100.0~GeV/c$^2$ DM mass and is negligible for the 1.0~GeV/c$^2$ DM mass. Thus, for higher DM masses, the contributions of both the break-up scenario and elastic scattering have to be considered when calculating the total scattering spectrum.

\section{Likelihood framework} \label{sec:likelihood}

In the standard scenario of the DM particle scattering off the detector nucleus, the scattering rate depends on two unknown DM parameters, the DM particle mass ($m_\chi$) and the spin-independent scattering cross-section of the DM particle with the detector nucleon ($\sigma_{\chi-SM}$). The functional form of the expected recoil spectrum of the darkonium scattering shows that the spectrum depends on another unknown parameter, $\gamma$~\cite{Laha_2014}. Thus, in total, there are three unknown parameters for this scenario, $\sigma_{\chi-SM}$, $\gamma$ and $\mu_n^2$, where $\mu_n^2$ is the reduced DM-nucleus mass. These three parameters are shown in Eq.~\ref{eq:darkonium_elastic_color} for the elastic scattering scenario: 

\begin{multline}\label{eq:darkonium_elastic_color}
    \left(\frac{d(\sigma v)}{dE_R}\right)_{A+2} = \frac{2m_A}{\pi v}\cdot \frac{\pi {\sigma_{\chi-SM}}A^2F_N^2(q)}{{\mu_n^2}} \\
    \cdot \left|\frac{4{\gamma}}{q}\text{tan}^{-1}\left(\frac{q}{4{\gamma}}\right)\right|^2\Theta(v-q/2{\mu_2}),
\end{multline}

where the details of the different parameters can be found in Ref.~\cite{Laha_2014}. In order to calculate the exclusion limits in the standard scenario for different $m_\chi$ using the profile likelihood framework, $m_\chi$ is fixed, and the value of the parameter of interest (POI), i.e. $\sigma_{\chi-SM}$, which fits the data best is found. This is then compared to the fit of the POI that gives the desired confidence level, using the likelihood ratio and the defined test statistic. The method is described well in Ref.~\cite{Cowan_2010}. However, this approach cannot directly be used in the darkonium scattering scenario as we have another POI. We use a different method to simplify and easily visualize the results. The DM mass $m_\chi$ is initially fixed to a particular value so that we are left with two parameters, one of which we can set a limit on. Now, a similar calculation can be performed by choosing different values of $\sigma_{\chi-SM}$ and fitting the spectrum to set the exclusion limits on the values of $\gamma$. The likelihood ratio in this approach takes the form:

\begin{equation}
    \lambda(\gamma) = \frac{\mathcal{L}(\gamma_{excl},\hat{\hat{\theta}}_b)}{\mathcal{L}(\gamma_{best},\hat{\theta}_b)}=\frac{\mathcal{L}_{excl}}{\mathcal{L}_{best}}
\end{equation}

where $\hat{\hat{\theta}}_b$ describes all the nuisance parameters of the background with a fixed $\gamma_{excl}$ that gives the desired confidence level, and $\hat{\theta}_b$ describes all the nuisance parameters of the background which fit the data best with $\gamma_{best}$ set free, for the probed $\sigma_{\chi-SM}$ (with a fixed $m_\chi$). Following this approach gives the exclusion limits in the $\gamma$ vs.  $\sigma_{\chi-SM}$  plane for that particular fixed mass $m_\chi$. This calculation can then be performed for various masses, and exclusion limits can be extracted for each $m_\chi$.

We conduct calculations across four distinct DM masses $m_\chi$ $\in \{0.5, 1.0, 10.0, 100.0 \}$ GeV/c$^2$. The data used comprises the first results from the CRESST-III DM search~\cite{DetA_2019} for $m_\chi$ = 0.5 and 1.0 GeV/c$^2$, and findings from XENON1T \cite{Xenon1T_2019} for $m_\chi$ = 10.0 and 100.0 GeV/c$^2$. In Ref. \cite{gupta_phd}, the calculations are shown for all four masses with only CRESST data. For each mass, we determine the minimum value of $\sigma_{\chi-SM}$ that the detector is sensitive to as the smallest probed $\sigma_{\chi-SM}$, establishing exclusion limits on $\gamma$. Subsequently, we derive further exclusion limits by incrementing $\sigma_{\chi-SM}$.

For the light masses, an unbinned likelihood function is constructed by fitting the data in the light yield vs. recoil energy plane. More information on the empirical form of the likelihood function can be found in Ref. \cite{CRESST_likelihood}. For heavy masses, we used a Poissonian binned likelihood approach where the used data, the background information, bin-width, and energy region of interest are as suggested by XENON1T \cite{Xenon1T_2019}. 

\section{Results and Conclusions}\label{sec:Conclusion}

\begin{figure}[h!]
\centering
\includegraphics[width=.48\textwidth, height=0.35\textwidth]{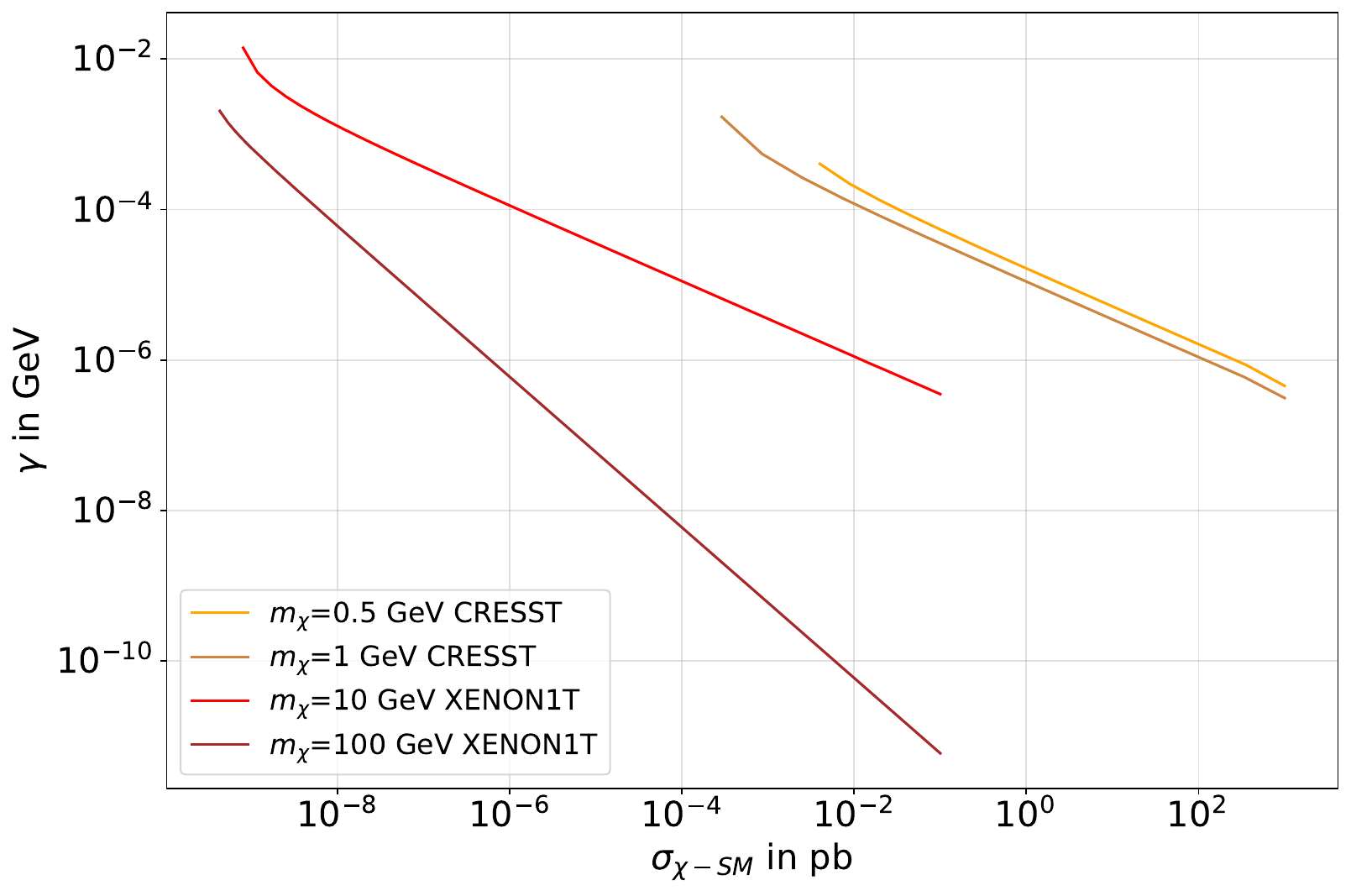}
\caption{The 90\% upper exclusion limits on the inverse scattering length $\gamma$ with respect to the dark matter particle-nucleon scattering cross-section, $\sigma_\text{{$\chi-SM$}}$ for $m_\chi$ = 0.5, 1.0, 10.0 and 100.0~GeV/c$^2$, using the data from the first results of the CRESST-III DM search~\cite{DetA_2019} for $m_\chi$ = 0.5 and 1.0 GeV/c$^2$, and from XENON1T~\cite{Xenon1T_2019} for $m_\chi$ = 10.0 and 100.0 GeV/c$^2$. }
\label{fig:limits_gamma}
\end{figure}

The 90\% confidence level upper limits on $\gamma$ are plotted in Fig.~\ref{fig:limits_gamma} for four different DM particle masses. It can be seen that the limits decrease with increasing value of $\sigma_{\chi-SM}$ as lowering $\gamma$ scales down the spectrum. Thus, increased $\sigma_{\chi-SM}$ is accommodated by lowering $\gamma$ to fit the observed spectrum. 

The upper exclusion limits on the value of $\gamma$ can be converted into lower exclusion limits on the $\sigma_{\chi-\chi}$ using Eq.~\ref{eq:sigma_gamma}. Since the constraints from astrophysics and N-body simulations are on the value of $\sigma_{\chi-\chi}/m_\chi$, the exclusion limits are translated to $\sigma_{\chi-\chi}/m_\chi$ and converted to cm$^2$/g.

As can be seen in Eq.~\ref{eq:sigma_gamma}, the value of $\sigma_{\chi-\chi}/m_\chi$ depends not only on $\gamma$ but also on the relative momentum between the DM particles. This opens up the ability to compare the exclusion to astrophysical observations at both the small-scale, with the typical velocities of $\mathcal{O}(10)$~km/s, and the cluster scales, with the typical velocities of $\mathcal{O}(1000)$~km/s. This velocity $\mathbf{v}$ should not be confused with the velocity $v$ in Eq.~\ref{eq:darkonium_elastic_color}. The former represents the relative velocity between the DM particles when the bound state is formed, whereas the latter represents the velocity of the bound states in the Milky Way with respect to the Earth. 

\begin{figure*}[t!]
\centering
\includegraphics[width=.48\textwidth, height=0.35\textwidth]{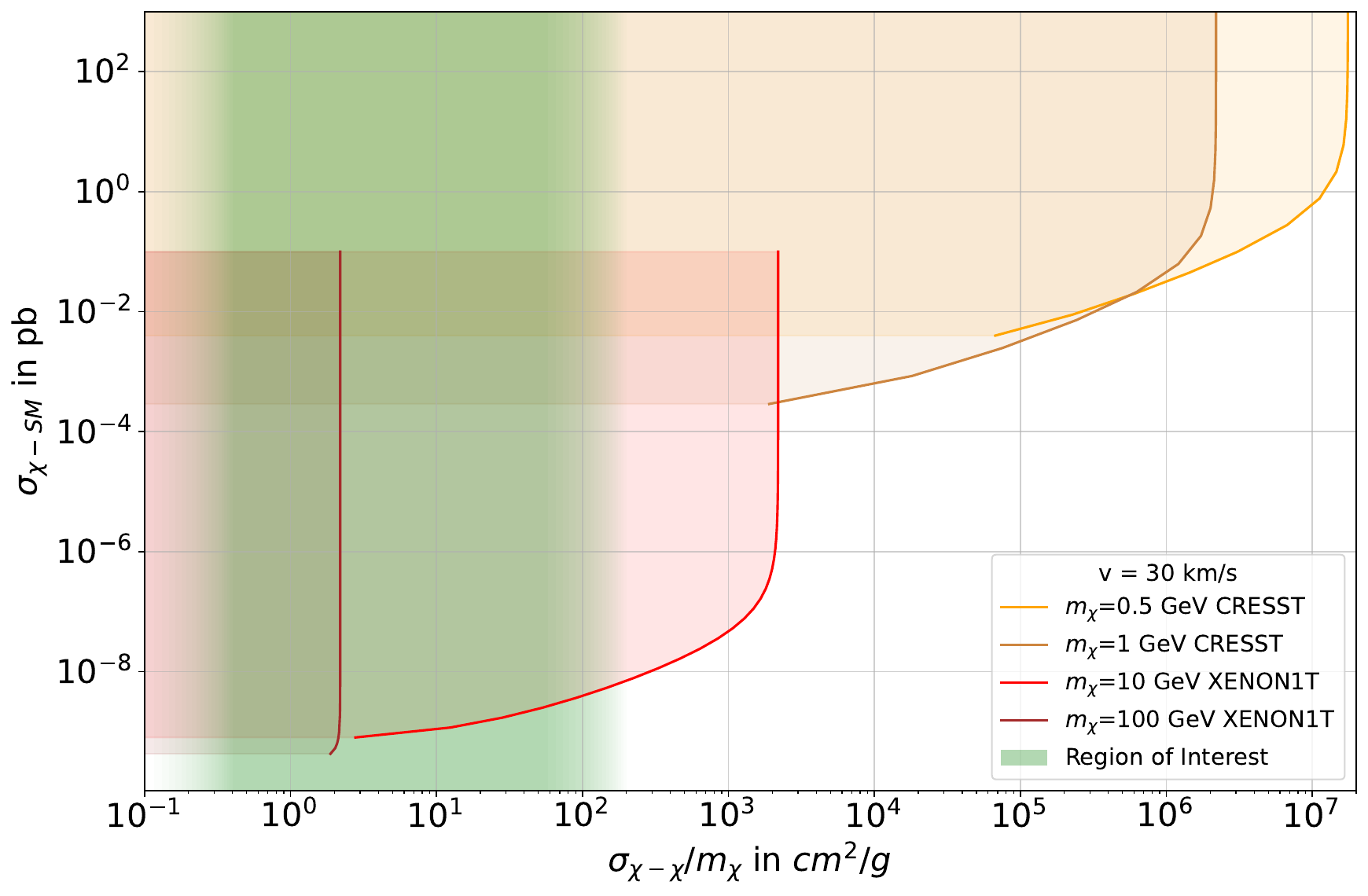}\hspace{2pt}
\includegraphics[width=.48\textwidth, height=0.35\textwidth]{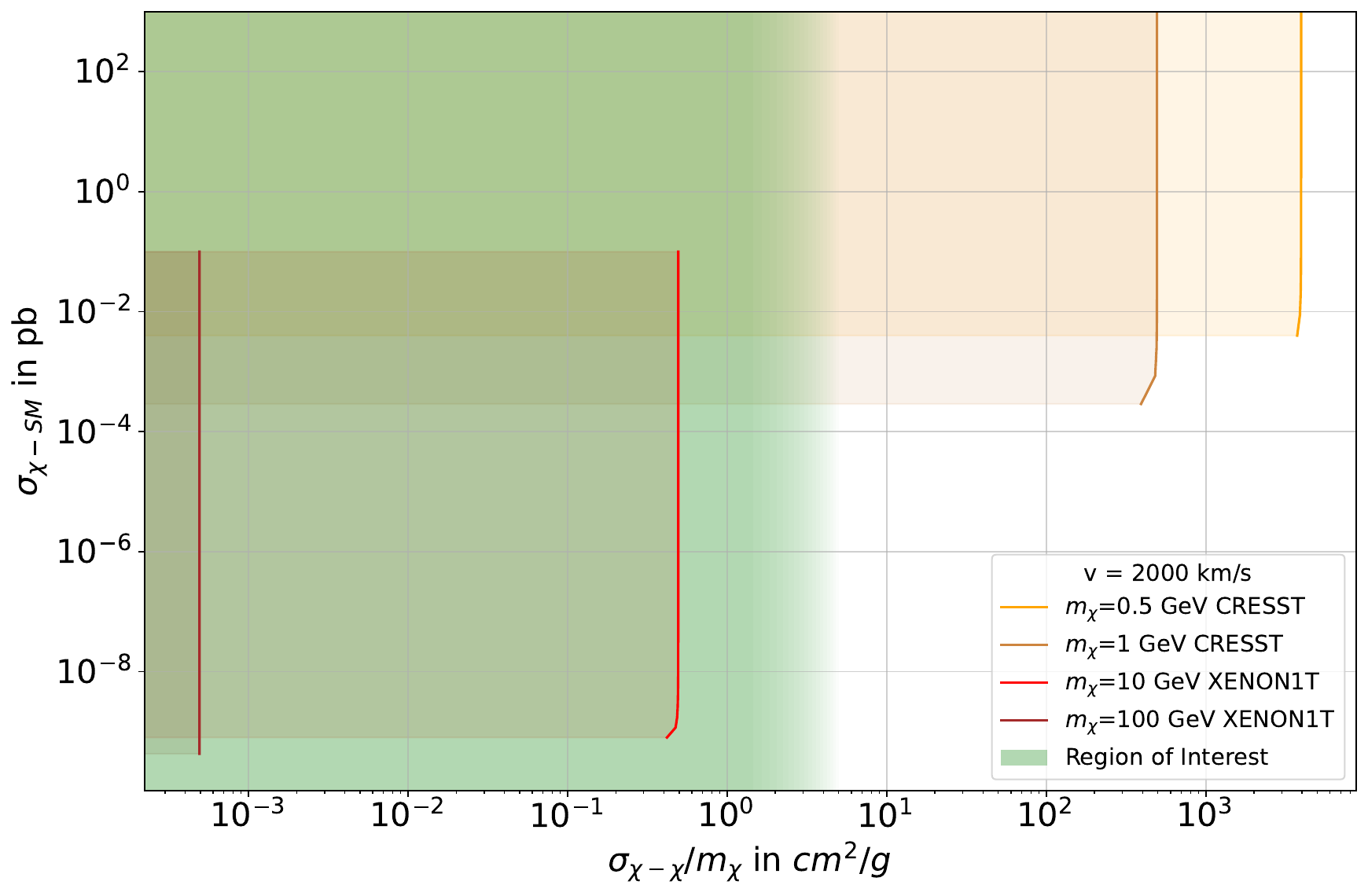}\\
\caption{90\% lower exclusion limits on the self-interaction cross-section $\sigma_{\chi-\chi}/m_\chi$ of DM particles at typical velocities $\mathbf{v}$ for different dark matter particle-nucleon scattering cross-sections $\sigma_{\chi-SM}$ calculated with the data from first results of the CRESST-III DM search~\cite{DetA_2019} for $m_\chi$ = 0.5 and 1.0 GeV/c$^2$, and from XENON1T~\cite{Xenon1T_2019} for $m_\chi$ = 10.0 and 100.0 GeV/c$^2$. The shaded area shows the excluded region for each mass at both velocity scales. The exclusion limits are compared with the preferred region of interest (ROI) for $\sigma_{\chi-\chi}/m_\chi$ (green band). This ROI comes from the constraints from astrophysics and N-body simulation at small-scales with $\mathbf{v}=30$~km/s (left)~\cite{Kahlhoefer_2019, Nishikawa_2020} and at cluster scales with $\mathbf{v}=2000$~km/s (right)~\cite{Kaplinghat_2016, Randall_2008,  Yoshida_2000, Rocha_2013}. The faded boundaries of the ROI acknowledge the potential influence of darkonium-darkonium self-interaction on the cross-section, which has not been considered in the calculation of the exclusion limits. The fade extends the cross-section boundaries to 4$\sigma_{\chi-\chi}/m_\chi$ (See Sec.~\ref{sec:Expected_spec} for a detailed discussion).}
\label{fig:limits_final}
\end{figure*}

The exclusion limits on $\gamma$ in Fig.~\ref{fig:limits_gamma} are calculated for the elastic scattering scenario only using $m_\chi$ = 0.5, 1.0 and 10.0 GeV/c$^2$, and the break-up scenario is neglected. For $m_\chi=100.0$~GeV/c$^2$, the calculation is done considering the total scattering scenario, including the break-up of the darkonium state. The exclusion limits on $\sigma_{\chi-\chi}/m_\chi$, shown in Fig.~\ref{fig:limits_final}, are calculated using a velocity of $\mathbf{v}=30$~km/s to compare them with the current constraints on small-scale structures from the astrophysical observations and simulations at $\sigma_{\chi-\chi}/m_\chi=(0.1-50)$~cm$^2$/g~\cite{Kahlhoefer_2019, Nishikawa_2020}. For comparison with the current constraints from cluster mergers at $\sigma_{\chi-\chi}/m_\chi<1.25$~cm$^2$/g, the exclusion limits are calculated at $\mathbf{v}=2000$~km/s~\cite{Kaplinghat_2016, Randall_2008}. As our predicted constraints return the estimate for $\sigma_{\chi-\chi}/m_\chi$, it is possible to match the astrophysical observations and our exclusion limits. 

We calculate exclusion limits for $\chi-\chi$ self-interactions only, while disregarding interactions between darkonium particles (see Sec. \ref{sec:Expected_spec}). Using the relation between $\gamma$ and $\sigma_{D-D}$ (Eq.~\ref{eq:D_D}) we convert the constraint on $\sigma_{D-D}$ from small-scale structures and merging clusters to constraints on $\sigma_{\chi-\chi}$. The direct detection exclusion limits on $\sigma_{\chi-\chi}/m_\chi$ for both the velocity scales are shown in Fig. ~\ref{fig:limits_final}, and the corresponding excluded regions are at the left (lower $\sigma_{\chi-\chi}/m_\chi$) of the exclusion limit for the given mass. The region of interest (in green) for both limits from astrophysical observations is depicted with faded boundaries, acknowledging the different composition (fermions or bosons) of the darkonium-bound state. 

For light DM particles ($m_\chi=0.5$ and 1~GeV/c$^2$), the $\sigma_{\chi-\chi}/m_\chi$ region of interest (ROI) is completely excluded at both velocity scales (the yellow and brown shaded regions) for the probed $\sigma_{\chi-SM}$. This is due to the darkonium form factor (Eq. 11 in \cite{Laha_2014}), where the exchanged momentum $q$ for lighter DM particles becomes comparable to $\gamma$, causing the form factor to approach its asymptotic value of 1. This means that if DM exists as darkonium, as proposed by \cite{Laha_2014}, with $\sigma_{\chi-\chi}/m_\chi$ in the ROI and a mass of 0.5 or 1.0~GeV/c$^2$, the effect of darkonium's internal structure on the recoil energy spectrum at CRESST cannot be observed for these masses. Therefore, the internal structure of darkonium can only be explored for heavier DM particles.

It can also be seen that there exists a lower limit to the sensitivity to $\sigma_{\chi-\chi}/m_\chi$ for any given mass as the exclusion limits remain constant at higher $\sigma_{\chi-SM}$ values. For example at 100~GeV/c$^2$, $\sigma_{\chi-\chi}/m_\chi>2.2$~cm$^2$/g cannot be probed at $v=30$~km/s. This occurs as an increasing $\sigma_{\chi-SM}$ decreases the value of $\gamma$ (Fig.~\ref{fig:limits_gamma}), and for very small values of $\gamma$ where $\gamma \ll k$, $\sigma_{\chi-\chi}/m_\chi$ depends only on $k$ (Eq.~\ref{eq:sigma_gamma}), which is constant for a given $\mathbf{v}$ and $m_\chi$. For cluster scale velocities, the limits are seen to be almost constant for all the $\sigma_{\chi-SM}$ probed due to the large value of $k$, whereas at small-scale velocities, this is seen only for high $\sigma_{\chi-SM}$ where $\gamma$ is small enough. Thus, probing higher $\sigma_{\chi-SM}$ cross-sections does not gain any sensitivity and only lower $\sigma_{\chi-SM}$ will allow us to probe more parameter space. 

In this study, we establish the first 90\% confidence level direct detection exclusion limits on the self-interaction cross-section of DM particles in the universal bound states as suggested by Laha and Braaten in Ref. \cite{Laha_2014}. The limits are calculated only for the scattering of darkonium off the detector nuclei for four different DM masses of $m_\chi$ = 0.5, 1.0, 10.0, and 100.0 GeV/c$^2$ using the data from CRESST-III and XENON1T. The exclusion limits are formulated in terms of the inverse scattering length. They are subsequently converted to the self-interaction cross-section using Eq. \ref{eq:sigma_gamma}, which also depends on the relative momentum between the DM particles. This methodology facilitates self-interaction comparisons across various velocity scales and, thus, can be directly juxtaposed with astrophysical and N-body simulation constraints on the self-interaction. For low-mass DM particles, the findings exclude the necessary self-interaction cross-sections required to address the small-scale crises and the allowed limit from cluster merger observations. However, this occurs within the scope of current probeable DM-nucleus scattering cross-sections, and exploring lower cross-sections could potentially unveil the required self-interacting cross-sections.

\begin{acknowledgements}
We especially thank Ranjan Laha, Eric Braaten, and Xiayong Chu for their detailed discussions on this model, which helped structure the final results and clarified many concepts. We are grateful to Laboratori Nazionali del Gran Sasso- INFN for their generous support of CRESST. This work has been funded by the Deutsche Forschungsgemeinschaft (DFG, German Research Foundation) under Germany’s Excellence Strategy-EXC 2094-390783311 and through the Sonderforschungsbereich (Collaborative Research Center) SFB-1258 ”Neutrinos and Dark Matter in Astro- and Particle Physics”, by BMBF Grants No. 05A20WO1 and No. 05A20VTA and by the Austrian science fund (FWF): Grants No. I5420-N, No. W1252N27, No. FG1, and by the Austrian research promotion agency (FFG) project No. ML4CPD. J. B. and H. K. were funded through the FWF project No. P 34778N ELOISE. The Bratislava group acknowledges a partial support provided by the Slovak Research and Development Agency (projects No. APVV-15-0576 and No. APVV-21-0377). The computational results presented were obtained using the CBE cluster of the Vienna BioCenter.
\end{acknowledgements}

%\renewcommand{\appendixname}{Appendix}
% \bibliographystyle{unsrtnat}
%\bibliographystyle{unsrtnat}
%\bibliography{main}
\printbibliography
\end{document}